\documentclass[english,prl, twocolumn]{revtex4-2}
\usepackage[T1]{fontenc}
\usepackage[utf8]{inputenc}
\usepackage{color}
\usepackage{babel}
\usepackage{verbatim}
\usepackage{bm}
\usepackage{amsmath}
\usepackage{amssymb}
\usepackage{graphicx}
\usepackage{float}
\usepackage[pdfusetitle,
 bookmarks=true,bookmarksnumbered=false,bookmarksopen=false,
 breaklinks=false,pdfborder={0 0 1},backref=false,colorlinks=true]
 {hyperref}
\hypersetup{
 colorlinks=true,linkcolor=blue,citecolor=blue,urlcolor=blue}

\makeatletter

\providecommand{\tabularnewline}{\\}

\ifdefined\showcaptionsetup
 \PassOptionsToPackage{caption=false}{subfig}
\fi
\usepackage{subfig}
\makeatother
\newcommand{\tcell}[1]{\begin{tabular}{@{}c@{}}#1\end{tabular}}

\begin{document}
\title{Geometric Field Theory for Elastohydrodynamics of Cosserat Rods}
\author{Mingjia Yan}
\email{my444@cam.ac.uk}

\thanks{Corresponding author. Authors share equal credit}
\affiliation{Department of Applied Mathematics and Theoretical Physics, Centre
for Mathematical Sciences, University of Cambridge, Cambridge CB3
0WA, United Kingdom}
\author{Mohamed Warda}
\email{mrmaw2@cam.ac.uk}

\thanks{Authors share equal credit}
\affiliation{Department of Applied Mathematics and Theoretical Physics, Centre
for Mathematical Sciences, University of Cambridge, Cambridge CB3
0WA, United Kingdom}
\author{Balázs Németh}
\email{bn273@cam.ac.uk}

\affiliation{Department of Applied Mathematics and Theoretical Physics, Centre
for Mathematical Sciences, University of Cambridge, Cambridge CB3
0WA, United Kingdom}
\author{Lukas Kikuchi}
\affiliation{Department of Applied Mathematics and Theoretical Physics, Centre
for Mathematical Sciences, University of Cambridge, Cambridge CB3
0WA, United Kingdom}
\author{Ronojoy Adhikari}
\email{ra413@cam.ac.uk}

\affiliation{Department of Applied Mathematics and Theoretical Physics, Centre
for Mathematical Sciences, University of Cambridge, Cambridge CB3
0WA, United Kingdom}
\begin{abstract}
Slender structures are ubiquitous in biological and physical systems,
from bacterial flagella to soft robotic arms. The Cosserat rod provides
a mathematical framework for slender bodies that can stretch, shear,
twist and bend. In viscous fluid environments at low Reynolds numbers---as
encountered in soft matter physics, biophysics, and soft continuum
robotics---inertial effects become negligible, and hydrodynamic forces
are well approximated by Stokes friction. We demonstrate that the
resulting elastohydrodynamic equations of motion, when formulated
using Cartan's method of moving frames, possess the structure of a
geometric field theory in which the configuration field takes values
in $SE(3)$, the Lie group of rigid body motions. This geometric formulation
yields coordinate-independent equations that are manifestly invariant
under spatial isometries and naturally suited to constitutive modeling
based on Curie's principle. We derive integrability conditions that
determine when constitutive laws can be derived from an energy functional,
thereby distinguishing between passive and active material responses.
We also obtain the beam limit for small deformations. Our results
establish a unified geometric framework for the nonlinear mechanics
of slender structures in slow viscous flow and enable efficient numerical
solutions. 
\end{abstract}
\maketitle

\section{Introduction}

Mathematical reformulations often reveal hidden structures in physical
theories. A good example is provided by the historical development
of Maxwell\textquoteright s equations, which progressed from their
original component representation through vector and tensor notation
to the succinct expression in terms of differential forms. Each stage
of reformulation preserved the underlying physics while uncovering
increasing levels of geometric insight \citep{Maxwell1891,Flanders1963,LandauLifshitz_fields,Jackson1975}.
Motivated by this perspective, we present a geometric field theory
for the elastohydrodynamics of slender elastic structures in slow
viscous flow as a reformulation of overdamped Cosserat rod theory. 

Cosserat rod theory \citep{Cosserat1896} generalizes classical beam
theory to incorporate stretching, shearing, twisting and bending in
a unified manner. It is broadly applicable, ranging from DNA mechanics
\citep{Balaeff-etal2006} and bacterial flagellar dynamics \citep{Ko-etal2017}
to soft robotics \citep{TillAloiRucker2019}. Mathematically, the
rod is described by a centerline and an orthonormal frame attached
to each cross-section of the rod. The deformation of the rod is described
in terms of spatial derivatives of the centerline and the frame. When
these derivatives at any cross-section are resolved in the local frame
at that cross-section, the components are invariant under rigid-body
motion. This yields intrinsic equations of motion and enables constitutive
modeling that respects the principle of material indifference.

It has long been recognized that the intrinsic equations of motion
for the conservative dynamics of Cosserat rods can be formulated in
terms of a Lie group valued configuration variable via Hamilton's
principle suitably adapted to exploit the group structure\citep{Arnold,MarsdenRatiu}.
The adaptation of Hamilton's principle to configuration variables
in homogeneous spaces approach traces back to Poincaré's 1901 paper
\citep{Poincare1901b} in which he expressed Euler-Lagrange equations
of motion in terms of Lie algebraic quantities. These equations are
now referred to variously as the Euler-Poincaré equations \citep{HoMaRa}.
The Hamiltonian character of these equations was later recognized
by Chetaev \citep{Chetaev1927,Chetaev1941}. The Poincaré-Chetaev
variational calculus has been used to derive the intrinsic equations
of motion for Cosserat rods from a Lagrangian that is invariant under
spatial isometries. 

This route to the intrinsic equations of motion is not available when
inertia is neglected and non-conservative forces are included. Here
we present an alternative route that avoids Poincaré-Chetaev variations
but takes as its point of departure the kinematic and dynamic equations
of Cosserat rod theory in vectorial form. By employing Cartan's method
of moving frames the vectorial equations are first presented in intrinsic
form and then their Lie group and differential form character are
uncovered. The resulting equations of motion have the character of
a geometric field theory for a matrix-valued field incorporating both
the positional and orientational degrees of freedom of the rod. The
elastohydrodynamic equations of motion separate into a kinematic part,
which is identical to Cartan frame equations. The kinematic compatibility
condition is recognized to be Cartan's equation of structure. The
dynamic part of the equations of motion acquires a form that is similar
to the sourced part of the Maxwell equations. Finally, we show these
equations can be expressed in terms of differential forms rather than
as partial differential equations. 

The paper is organized as follows. In Section \ref{sec:Cosserat-rods},
we present Cosserat rod theory and derive the governing equations
in vectorial form. Section \ref{sec:moving-frames} reformulates these
equations using Cartan's method of moving frames \citep{Cartan_RG}
revealing their invariant structure. In Section \ref{sec:Geometric-field-theory},
a geometric field theory is established on $SE(3)$, unifying kinematics
and dynamics. Then, Section \ref{sec:Constitutive-laws} derives the
general constitutive laws via a work-energy principle as integrability
conditions on stress and external force, yielding the corresponding
boundary conditions after integration by parts. Following immediately
is the linearization presented in Section \ref{sec:Linearisation}.
In Section \ref{sec:Coordinatised-equations}, we derive explicit
coordinate equations for planar Cosserat rod motions for both the
general and hyperelastic linear constitutive laws. In Section \ref{sec:Beam-limits},
we recover well-known beam equations from the linearized coordinatized
equations under the hyperelastic constitutive law as a limiting case.
Section \ref{sec:Discussion-and-Conclusion} concludes with discussion
of results and future directions. 

\section{\label{sec:Cosserat-rods}Cosserat rods}

\begin{figure}[h]
\centering{}

\includegraphics[scale=0.4]{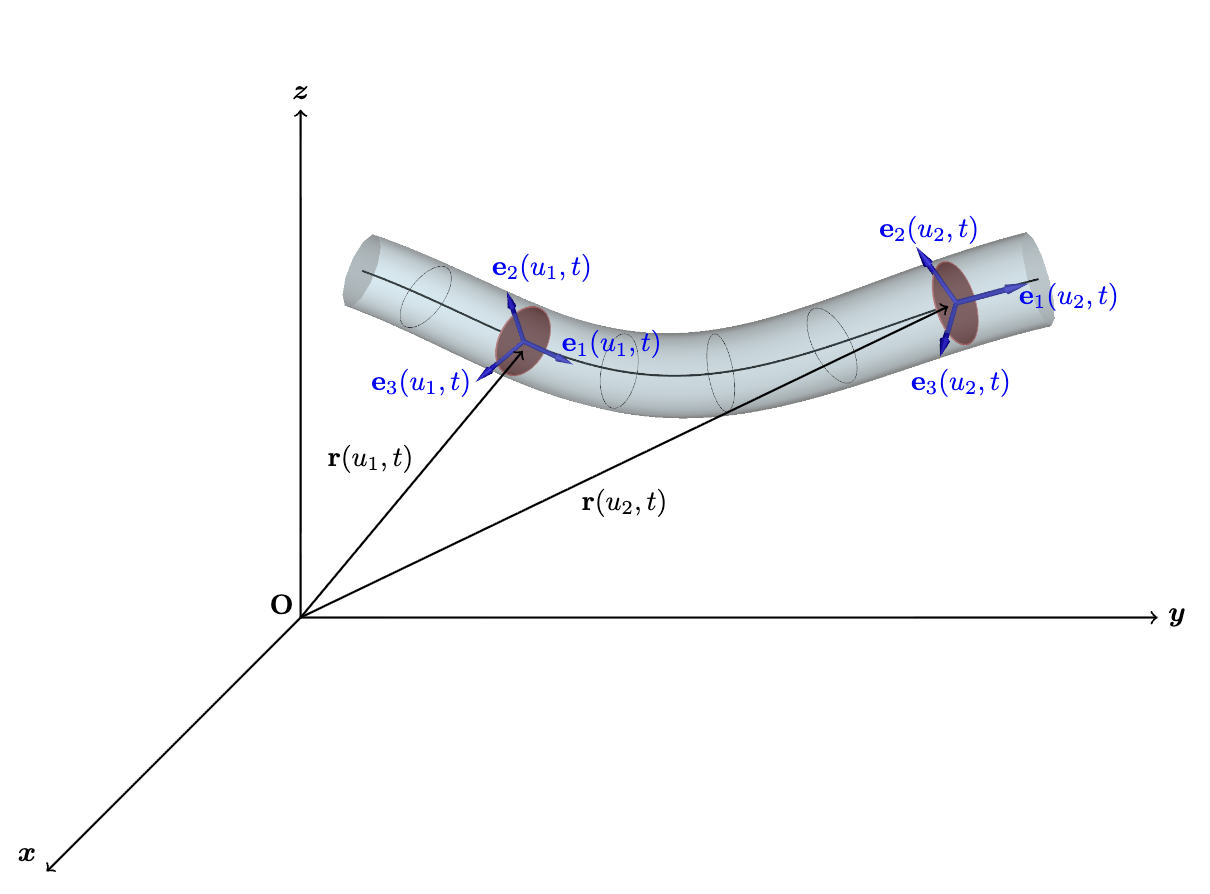}

\caption{Kinematics of the Cosserat rod showing the centerline $\boldsymbol{r}(u,t)$
and the orthonormal frames $\boldsymbol{e}_{i}(u,t)$ rigidly attached
to the cross-section at material parameter $u$ at time $t$. \label{fig:Cosserat-rod-configurations}}
\end{figure}

The Cosserat rod is an elastic rod model that can stretch, shear,
twist and bend, capturing all possible deformation modes of a slender
rod \citep{Antman}. The configuration of a Cosserat rod at time $t$
can be described by the four vector-valued functions, $\boldsymbol{r}(u,t),\boldsymbol{e}_{1}(u,t),\boldsymbol{e}_{2}(u,t),\boldsymbol{e}_{3}(u,t),$
where $\boldsymbol{r}(u,t)$ is the centerline position and $\boldsymbol{e}_{1}(u,t),\boldsymbol{e}_{2}(u,t),\boldsymbol{e}_{3}(u,t)$
forms an orthonormal triad rigidly attached to each cross-section,
yielding a moving frame. Here, $u\in[0,L]$ is the material parameter
identifying cross-sections. Thus, six degrees of freedom, three for
centerline position and three for frame orientation, completely characterize
the rod's configuration at any time. We choose $\boldsymbol{e}_{1}$
normal to the cross-section and $\boldsymbol{e}_{2}$ and $\boldsymbol{e}_{3}$
spanning its plane, see Fig \ref{fig:Cosserat-rod-configurations}.

\textit{Kinematics.} The spatial and temporal partial derivatives
of the centerline and the frame vectors are given by
\begin{align}
\boldsymbol{r}^{\prime} & =\boldsymbol{h},\quad\boldsymbol{e}_{i}^{\prime}=\boldsymbol{\Pi}\times\boldsymbol{e}_{i},\label{eq:spatial-kinematic}\\
\dot{\boldsymbol{r}} & =\boldsymbol{v},\quad\dot{\boldsymbol{e}_{i}}=\boldsymbol{\Omega}\times\boldsymbol{e}_{i},\label{eq:temporal-kinematic}
\end{align}
where primes and dots denote partial derivatives with respect to $u$
and $t$ respectively and $i=1,2,3$. The variables $\boldsymbol{h}$
and $\boldsymbol{\Pi}$ are the positional and orientational deformations
while $\boldsymbol{v}$ and $\boldsymbol{\Omega}$ are the velocity
of the centerline and the angular velocity of the moving frame. The
square arc-length $ds^{2}=d\boldsymbol{r}\cdot d\boldsymbol{r}$ then
becomes $ds^{2}=\boldsymbol{r}^{\prime}\cdot\boldsymbol{r}^{\prime}du^{2}=h^{2}du^{2}$,
identifying the norm of $\boldsymbol{h}$  as the metric of the centerline.
We shall call Eq. (\ref{eq:spatial-kinematic}) and Eq. (\ref{eq:temporal-kinematic})
the \emph{spatial} and\emph{ temporal} kinematic equations, respectively.
These kinematic equations maintain the orthonormality of the frame
for all $u$ and $t$. The equality of mixed partial derivatives of
the centerline and frame yields the following \textit{compatibility
conditions} 
\begin{equation}
\dot{\boldsymbol{h}}-\boldsymbol{v}^{\prime}=0,\quad\dot{\boldsymbol{\Pi}}-\boldsymbol{\Omega}^{\prime}=\boldsymbol{\Pi}\times\boldsymbol{\Omega},\label{eq:compatibility-conditions}
\end{equation}
relating the deformations and the velocities. Together with the kinematic
equations $(\ref{eq:spatial-kinematic})$ and $(\ref{eq:temporal-kinematic})$,
the above three equations completely specify the kinematics of the
rod.

\textit{Dynamics.} The dynamics of the Cosserat rod follows from linear
and angular momentum balance, yielding
\begin{align}
\begin{split}\dot{\boldsymbol{p}} & =\boldsymbol{F}^{\prime}+\boldsymbol{f},\\
\dot{\boldsymbol{L}} & =\boldsymbol{M}^{\prime}+\boldsymbol{h}\times\boldsymbol{F}+\boldsymbol{m},
\end{split}
\end{align}
where $\boldsymbol{p}$ and $\boldsymbol{L}$ are the linear and angular
momentum densities, $\boldsymbol{F}$ and $\boldsymbol{M}$ are the
stress and moment stress on the cross-section while $\boldsymbol{f}$
and $\boldsymbol{m}$ are the body force and torque densities.

In the overdamped limit, inertial terms vanish, giving the balance
laws
\begin{equation}
\begin{split}\boldsymbol{F}^{\prime}+\boldsymbol{f} & =0,\\
\boldsymbol{M}^{\prime}+\boldsymbol{h}\times\boldsymbol{F}+\boldsymbol{m} & =0,
\end{split}
\label{eq:dynamics}
\end{equation}
The Cosserat stresses and body force and torque densities follow constitutive
laws relating them to the positional and orientational deformations.
This closes the system, giving six coupled PDEs for the centerline
position $\boldsymbol{r}(u,t)$ and frame vectors $\boldsymbol{e}_{i}(u,t)$,
supplemented by appropriate boundary conditions. We focus exclusively
on these overdamped equations and show how they can be geometrized
using Cartan's method of moving frames.

\section{\label{sec:moving-frames}moving frames}

The principle of material indifference requires constitutive laws
to be invariant under rigid transformations \citet{Antman}. This
invariance is most easily imposed by expressing the kinematic and
dynamic equations in the moving frame $\boldsymbol{e}_{i}$ and, following
Cartan, recognizing their Lie group and differential form character.
The former identifies the invariant components of the kinematic and
dynamic variables and enables them to be brought into invariant constitutive
relationship. The latter reveals the equations of motion to be a geometric
field theory and enables their structure-preserving numerical integration
using Lie group methods. 

The key idea of Cartan's method is to express vectors in the \textit{moving
frames} and express their rates of change in the same frame. For an
arbitrary vector $\boldsymbol{A}(u,t)$ in the moving frame, we can
write 
\begin{equation}
\boldsymbol{A}=\boldsymbol{e}_{i}A_{i},\quad\boldsymbol{e}_{i}\cdot\boldsymbol{A}=A_{i}
\end{equation}
The moving-frame component $A_{i}$ is invariant under global translations
and rotations. It is convenient to introduce the notation $\underline{A}=(A_{1},A_{2},A_{3})$
to denote the triple of moving frame components of a vector. The orthogonality
of the frames defines a dot and cross product for the moving frame
components, 
\begin{align}
\boldsymbol{A}\cdot\boldsymbol{B} & =A_{i}B_{i}=\underline{A}\cdot\underline{B}\\
\boldsymbol{e}_{i}\cdot(\boldsymbol{A}\times\boldsymbol{B}) & =\epsilon_{ijk}A_{j}B_{k}=(\underline{A}\times\underline{B})_{i}
\end{align}
The spatial and temporal rates of change of $\underline{A}$ resolved
in the moving frame take the form
\begin{align}
\begin{split}\boldsymbol{e}_{i}\cdot\boldsymbol{A}^{\prime} & =A_{i}^{\prime}+\epsilon_{ijk}\Pi_{j}A_{k}=(\underline{A}^{\prime}+\underline{\Pi}\times\underline{A})_{i}\\
\boldsymbol{e}_{i}\cdot\boldsymbol{\dot{A}} & =\dot{A_{i}}+\epsilon_{ijk}\Omega_{j}A_{k}=(\underline{\dot{A}}+\underline{\Omega}\times\underline{A})_{i}.
\end{split}
\label{eq:derivatives-moving-frame}
\end{align}

These co-rotational derivatives represent the infinitesimal change
of the vector components in the moving frame taking into account the
rate of change of the frame itself. The terms involving the cross
product are ``Coriolis'' contributions due to the rotation of the
frame.

\emph{Kinematics}.\textbf{ }We express deformations and velocities
in the moving frame in the same fashion as 
\begin{align}
\boldsymbol{h} & =\boldsymbol{e}_{i}h_{i},\quad\boldsymbol{\Pi}=\boldsymbol{e}_{i}\Pi_{i}\\
\boldsymbol{v} & =\boldsymbol{e}_{i}v_{i},\quad\boldsymbol{\Omega}=\boldsymbol{e}_{i}\Omega_{i}.
\end{align}
where repeated indices are summed over. We can then derive the corresponding
kinematic equations in the moving frame by taking dot products following
Eq. $(\ref{eq:derivatives-moving-frame})$
\begin{align}
\boldsymbol{r}^{\prime} & =h_{i}\boldsymbol{e}_{i},\quad\boldsymbol{e}_{i}^{\prime}=\epsilon_{ijk}\Pi_{j}\boldsymbol{e}_{k}\\
\dot{\boldsymbol{r}} & =v_{i}\boldsymbol{e}_{i},\quad\dot{\boldsymbol{e}_{i}}=\epsilon_{ijk}\Omega_{j}\boldsymbol{e}_{k}.
\end{align}
Their moving frame components,
\begin{align}
h_{i} & =\boldsymbol{e}_{i}\cdot\boldsymbol{r}^{\prime},\quad\Pi_{i}=\epsilon_{ijk}\boldsymbol{e}_{j}\cdot\boldsymbol{e}_{k}^{\prime}\\
v_{i} & =\boldsymbol{e}_{i}\cdot\dot{\boldsymbol{r}},\quad\Omega_{i}=\epsilon_{ijk}\boldsymbol{e}_{j}\cdot\dot{\boldsymbol{e}}_{k},
\end{align}
are invariant under rigid motions $\boldsymbol{r}\rightarrow\boldsymbol{R}\cdot\boldsymbol{r}+\boldsymbol{a}$
and $\boldsymbol{e}_{i}\rightarrow\boldsymbol{R}\cdot\boldsymbol{e}_{i}$,
where $\boldsymbol{a}$ is a translation and $\boldsymbol{R}$ is
a $3\times3$ orthogonal rotation matrix. The invariance of the metric
\begin{equation}
g=\|\boldsymbol{h}\|=\sqrt{h_{1}^{2}+h_{2}^{2}+h_{3}^{2}}
\end{equation}
under those rigid transformations follows immediately. Therefore,
the invariants $h_{i}$ and $\Pi_{i}$ determine the configuration
up to a rigid motion. Conversely, two configurations with identical-valued
invariants differ from each other by a rigid motion. From their definitions,
the deformations and velocities in the moving frame are \emph{non-holonomic},
in the sense that they are not derivatives of functions of the coordinates.
The compatibility conditions are obtained by taking dot products on
both sides and using the relationship between derivatives in the fixed
and moving frames. The first compatibility condition gives
\begin{equation}
\dot{\underline{h}}+\underline{\Omega}\times\underline{h}=\underline{v}^{\prime}+\underline{\Pi}\times\underline{v}
\end{equation}
Similarly, the second compatibility condition gives 
\begin{equation}
\dot{\underline{\Pi}}-\underline{\Omega}^{\prime}=\underline{\Pi}\times\underline{\Omega},
\end{equation}

\emph{Dynamics}. In a similar manner, using the dynamic variables
in the moving frame \citep{Antman}
\begin{align}
\boldsymbol{F}=F_{i}\boldsymbol{e}_{i},\quad\boldsymbol{f}=f_{i}\boldsymbol{e}_{i},\quad & \boldsymbol{M}=M_{i}\boldsymbol{e}_{i},\quad\boldsymbol{m}=m_{i}\boldsymbol{e}_{i},
\end{align}
taking the dot product of the dynamical equations and using the relation
between derivatives in the fixed and the moving frames, we obtain
\begin{align}
\begin{split}\underline{F}^{\prime}+\underline{\Pi}\times\underline{F}+\underline{f} & =0\\
\underline{M}^{\prime}+\underline{\Pi}\times\underline{M}+\underline{h}\times\underline{F}+\underline{m} & =0
\end{split}
\end{align}
where the ``Coriolis'' terms involving cross products with $\underline{\Pi}$
appear due to the motion of the frame. The kinematic and dynamic equations
in the moving frame are invariant under rigid motions and a constitutive
relation between the dynamic and kinematic variables automatically
respect spatial isometries. The moving frame quantities, then, are
naturally suited for constitutive modeling. 

\section{\label{sec:Geometric-field-theory}Geometric field theory}

While the moving frame nature of Cosserat rod equations has been recognized
and exploited previously (see, e.g. \citep{Antman} for a detailed
treatment), a more geometrically consistent approach emerges by examining
the underlying geometric structure of Cosserat rod theory. This perspective
allows us to formulate the kinematics and dynamics using Lie groups
in a coordinate-independent way. To the best of our knowledge, the
Lie group and differential form perspective has not been developed
consistently for the case of a Cosserat rod in slow viscous flow.
We now present our development below.

\subsection{Lie groups}

\emph{Kinematics}. The position of the moving frame along the rod
is represented by the centerline vector $\boldsymbol{r}$, and the
orientation of the frame is represented by the three orthonormal vectors
$\boldsymbol{e}_{i},i=1,2,3$. Hence, the rod configuration combines
the centerline position $\boldsymbol{r}$ and frame orientation $\boldsymbol{e}_{i}$
into the following $4\times4$ matrix $\varphi\in SE(3)$, where $SE(3)$
is the special Euclidean group of rigid-body motions in 3D,
\begin{align}
\varphi & =\left[\begin{array}{cccc}
1 & 0 & 0 & 0\\
\boldsymbol{r} & \boldsymbol{e}_{1} & \boldsymbol{e}_{2} & \boldsymbol{e}_{3}
\end{array}\right].
\end{align}
For this to hold, we need to choose a fixed reference frame. Here,
$\boldsymbol{r}$ is a $3\times1$ vector and $\boldsymbol{e}_{1},\boldsymbol{e}_{2},\boldsymbol{e}_{3}$
form a $3\times3$ orthogonal rotation matrix giving, respectively,
the components of the centerline and the moving frame with respect
to the fixed frame. As a consequence, taking the spatial and temporal
partial derivatives gives the Lie-group kinematic equations concisely,
in an invariant manner, as
\begin{equation}
\varphi^{\prime}=\varphi E,\quad\dot{\varphi}=\varphi V\label{eq:kinematics-group}
\end{equation}
where the Lie algebra elements $E,V\in\mathfrak{se}(3)$ are the $4\times4$
matrices, given by
\begin{align}
E=\begin{bmatrix}0 & 0\\
\underline{h} & \hat{\Pi}
\end{bmatrix}\quad & V=\begin{bmatrix}0 & 0\\
\underline{v} & \hat{\Omega}
\end{bmatrix}
\end{align}
representing, respectively, the generalized deformation and the generalized
velocity. The $3\times3$ hatted matrices have components $\hat{A}_{ij}=\epsilon_{ijk}A_{k}$
for a vector $\underline{A}$ in the moving frame. From their definition,
it is clear that $E$ and $V$ are invariant under a global rigid
transformation $\varphi\rightarrow g\varphi$ where $g$ is a constant
matrix in $SE(3).$ 
\begin{figure*}[t]
\begin{minipage}[t]{0.35\textwidth}%
\subfloat[An undeformed Cosserat rod.]{\includegraphics[scale=0.45]{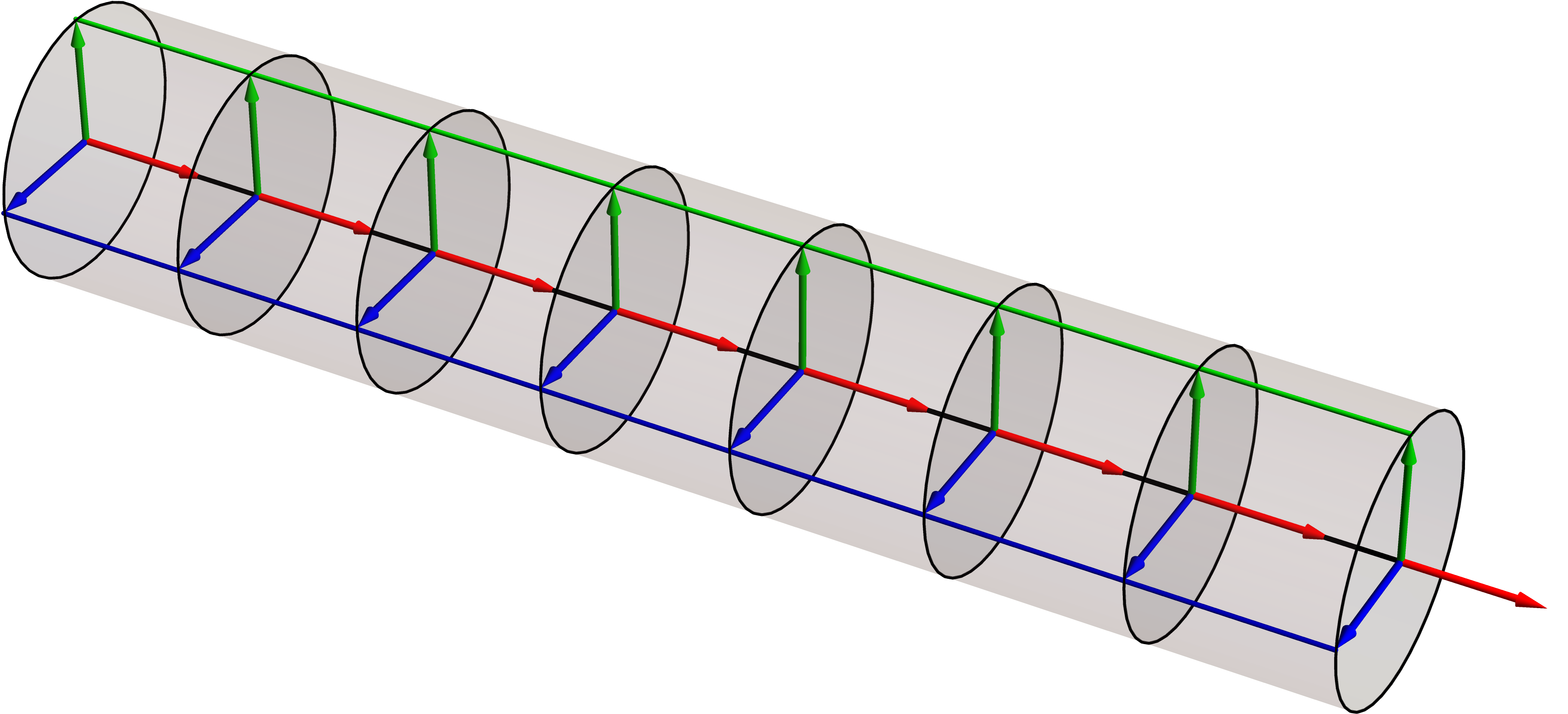}

}%
\end{minipage}%
\begin{minipage}[t]{0.35\textwidth}%
\subfloat[A stretched Cosserat rod.]{\includegraphics[scale=0.45]{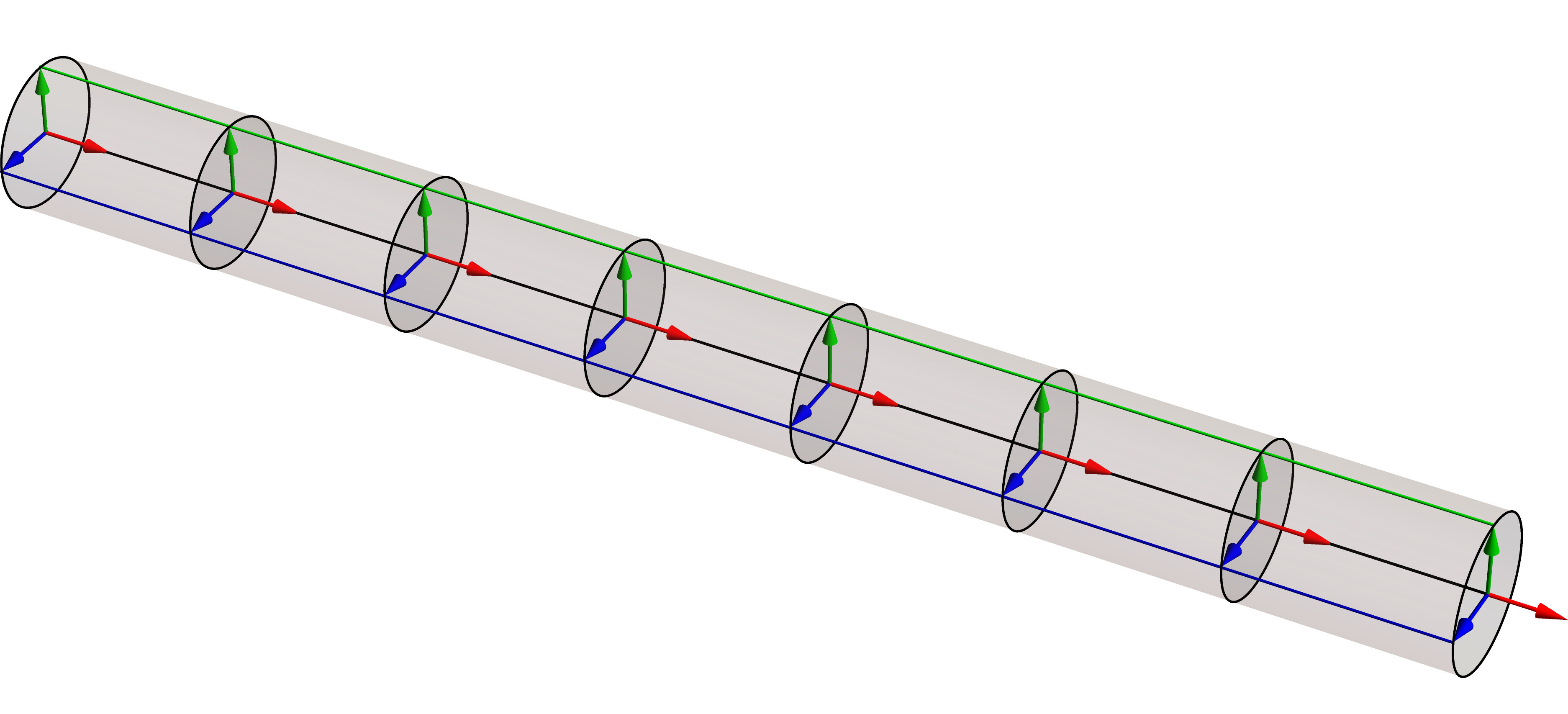}

}%
\end{minipage}%
\begin{minipage}[t]{0.35\textwidth}%
\subfloat[A sheared Cosserat rod.]{\includegraphics[scale=0.45]{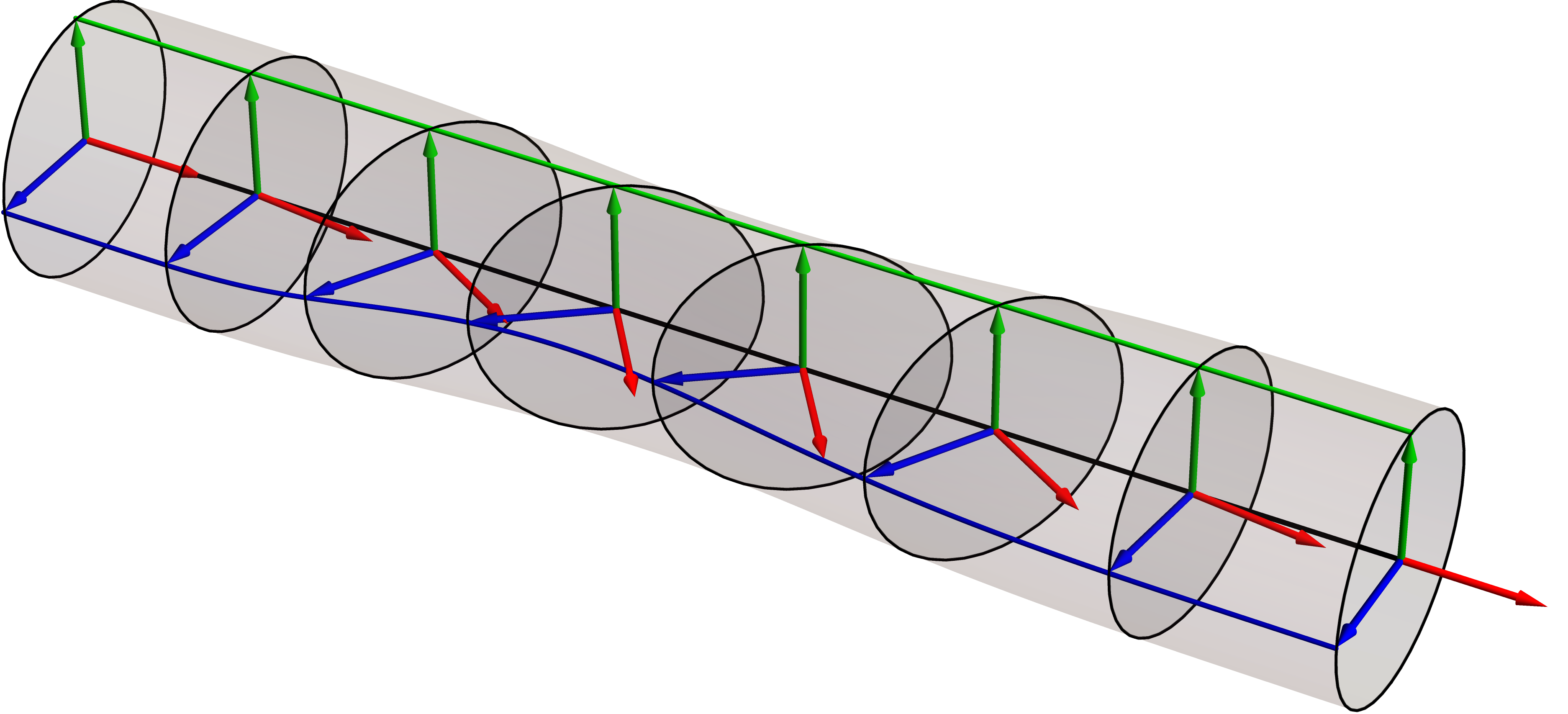}

}%
\end{minipage}

\begin{minipage}[t]{0.35\textwidth}%
\subfloat[A twisted Cosserat rod.]{\includegraphics[scale=0.45]{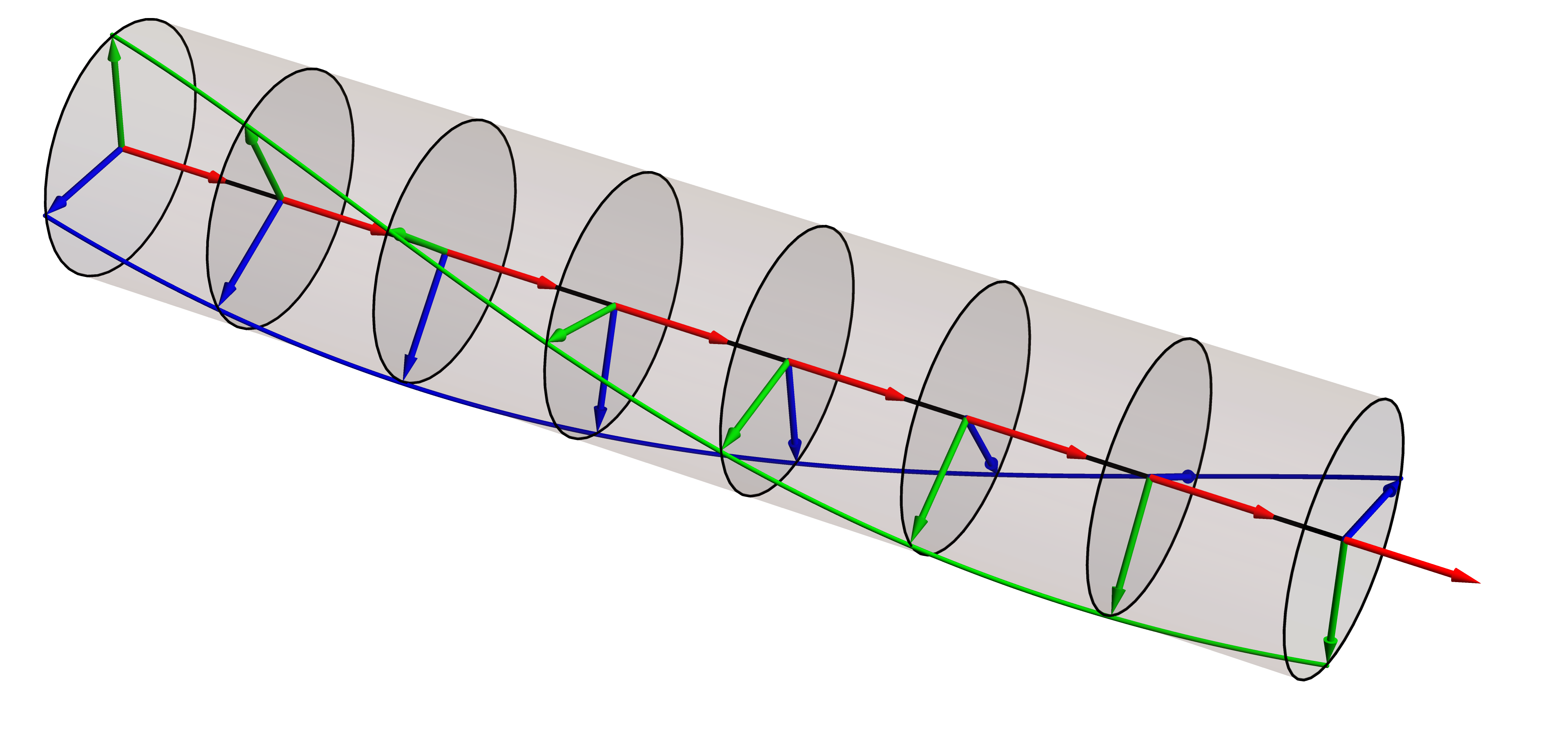}

}%
\end{minipage}%
\begin{minipage}[t]{0.35\textwidth}%
\subfloat[A bent Cosserat rod.]{\includegraphics[scale=0.45]{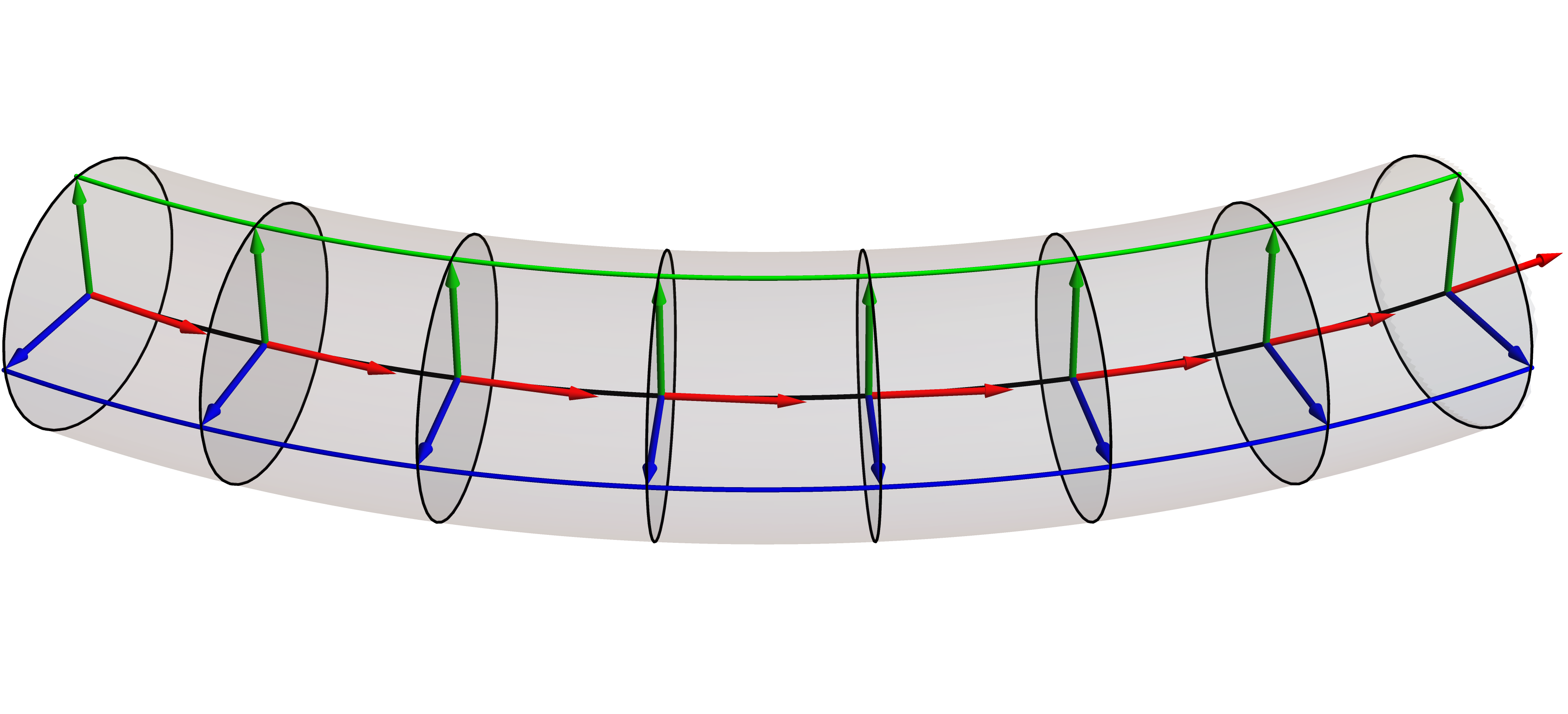}

}%
\end{minipage}%
\begin{minipage}[t]{0.35\textwidth}%
\subfloat[A sheared, twisted and bent Cosserat rod.]{\includegraphics[scale=0.45]{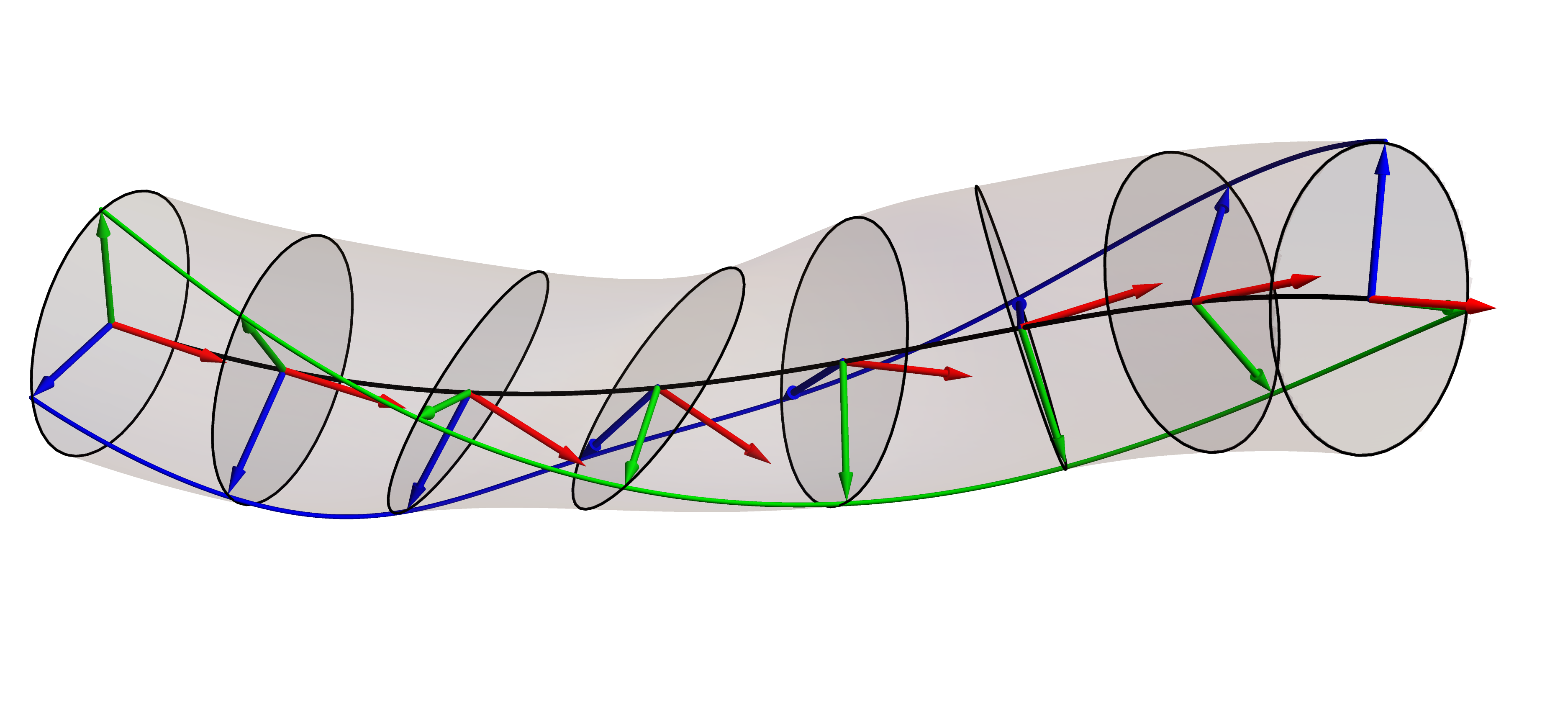}

}%
\end{minipage}

\caption{Examples of deformations of the Cosserat rod. (a) A straight Cosserat
rod with no deformation. (b-f) Stretching, shearing, twisting, bending
deformation modes, as well as the combined deformation mode. The transparent
light blue body is the bulk of the Cosserat rod, with the centerline
(black line) going through its radial centre. The material frames
and cross-sections are displayed at some intermediate points along
the centreline, where the frame vectors are $\boldsymbol{e}_{1}$
(red), $\boldsymbol{e}_{2}$ (blue), and $\boldsymbol{e}_{3}$ (green).
\label{fig:Cosserat-rod-deformations}}
\end{figure*}

The Lie algebra $\mathfrak{se}(3)$ is a $6$-dimensional vector space
with a basis $b_{\alpha}$ of six $4\times4$ matrices, where $\alpha=1,\ldots6.$
We can then write $E=E_{\alpha}b_{\alpha}$ and $V=V_{\alpha}b_{\alpha}$
to identify 
\begin{align}
E_{\alpha} & \longleftrightarrow\underline{E}=(h_{1},h_{2},h_{3},\Pi_{1},\Pi_{2},\Pi_{3}),\label{eq:E}\\
V_{\alpha} & \longleftrightarrow\underline{V}=(v_{1},v_{2},v_{3},\Omega_{1},\Omega_{2},\Omega_{3}).
\end{align}
Following screw theory \citep{Ball1876}, we call the pair of velocities
and angular velocities $(\underline{v},\underline{\Omega})$ the \textit{twist}
and the pair of positional and orientational deformations $(\underline{h},\underline{\Pi})$,
the \textit{turn}. The components of $E$ are associated with different
deformation modes: $h_{1}$ with stretching, $h_{2}$ and $h_{3}$
with shearing, $\Pi_{1}$ with twisting, and $\Pi_{2}$ and $\Pi_{3}$
with bending. See Fig \ref{fig:Cosserat-rod-deformations} for a visualization
of different deformation modes. The Lie algebra $\mathfrak{se}(3)$
is equipped with a Lie bracket $[\cdot,\cdot]$ defined as a matrix
commutator $[X,Y]=XY-YX$.The Lie algebra basis satisfies the following
Lie bracket relation \citep{Lee_manifold}
\begin{align}
[b_{\alpha},b_{\beta}] & =C_{\alpha\beta\gamma}b_{\gamma},\quad\alpha,\beta,\gamma=1,\ldots,6\label{eq:Lie-bracket}
\end{align}
where $C_{\alpha\beta\gamma}$ are the structure constants. The compatibility
condition from equality of mixed partial derivatives gives
\begin{equation}
\dot{E}-V^{\prime}=[E,V]\label{eq:compatibility-condition-group}
\end{equation}
where $[E,V]$ is the matrix commutator. In components, it reads
\begin{equation}
\dot{E}_{\alpha}-V_{\alpha}^{\prime}=C_{\alpha\beta\gamma}E_{\beta}V_{\gamma}
\end{equation}
where the structure constants $C_{\alpha\beta\gamma}$ are defined
by the Lie bracket $(\ref{eq:Lie-bracket})$ above. We note that the
matrix commutator is identical to the adjoint action $\text{ad}_{X}Y=[X,Y]$
of the Lie algebra on itself. The map $\mathrm{\mathrm{ad}}:\mathfrak{se}(3)\rightarrow\mathfrak{se}(3)$
has the matrix representation 
\begin{align}
\mathrm{ad}_{E} & =\begin{bmatrix}\hat{\Pi} & \hat{h}\\
0 & \hat{\Pi}
\end{bmatrix}
\end{align}
 in the Lie algebra basis. 

\subsection{Differential forms }

Via the matrix-valued differential forms 
\begin{equation}
d\varphi=\varphi^{\prime}du+\dot{\varphi}dt,\quad\xi=Edu+Vdt
\end{equation}
the kinematic equations $(\ref{eq:kinematics-group})$ and the compatibility
condition $(\ref{eq:compatibility-condition-group})$ can be written
compactly into
\begin{equation}
d\varphi=\varphi\xi,\quad d\xi+\xi\wedge\xi=0,\label{eq:cartan-structure}
\end{equation}
where the wedge product of a matrix-valued form is defined as $(\xi\wedge\xi)_{\alpha\beta}=\xi_{\alpha\gamma}\wedge\xi_{\gamma\beta}$.
The kinematic equations appear as the left-trivialization of the tangent
space of a Lie group to its Lie algebra. We note that $\varphi^{-1}d\varphi$
is called the \textit{Darboux derivative} of $\varphi$ in the differential
geometry literature, see, e.g. Griffiths \citep{Griffiths} for a
discussion. The compatibility condition represents Cartan's structure
equation for the pullback $\xi$ of the Lie algebra-valued left-invariant
Maurer-Cartan \citep{Cartan1904} form on $SE(3)$. 

\emph{Dynamics}. Stresses are dual to deformations, hence to define
the generalized stress properly as a dual Lie algebra $\mathfrak{se}^{*}(3)$
element we need to first clarify what the dual Lie algebra is as a
dual vector space. For a Lie algebra $\mathfrak{g}$, its dual, $\mathfrak{g}^{*}$,
is by definition the set of all linear maps from $\mathfrak{g}$ to
$\mathbb{R}$. Then, given a vector space basis $(b_{1},\dots,b_{n})$
of the Lie algebra $\mathfrak{g}$, the associated dual basis of $\mathfrak{g}$
is defined as a linear map $b_{\alpha}^{*}:\mathfrak{g}\rightarrow\mathbb{R}$
on basis elements by
\begin{align}
\langle b_{\alpha}^{*},b_{\beta}\rangle_{\mathfrak{g^{*}\times\mathfrak{g}}}:=b_{\alpha}^{*}\left(b_{\beta}\right)=\begin{cases}
1 & \alpha=\beta\\
0 & \alpha\neq\beta.
\end{cases}\label{eq:duality}
\end{align}
which forms a dual basis $(b_{1}^{*},\dots,b_{n}^{*})$ for $\mathfrak{g}^{*}$.
In this way, we may combine the stress and moment stress into a generalized
stress $\Sigma=\Sigma_{\alpha}b_{\alpha}^{*}$ in the dual Lie algebra
$\mathfrak{se^{*}}(3)$ with dual basis $b_{\alpha}^{*}$ and identify
it with a $6$-dimensional dual vector $\underline{\Sigma}$, as
\begin{equation}
\Sigma_{\alpha}\longleftrightarrow\underline{\Sigma}=(F_{1},F_{2},F_{3},M_{1},M_{2},M_{3}).
\end{equation}
The force and torque are combined into a generalized force $j=j_{\alpha}b_{\alpha}^{*}$
identified with a $6$-dimensional dual vector $\underline{j}$
\begin{equation}
j_{\alpha}\longleftrightarrow\underline{j}=(f_{1},f_{2},f_{3},m_{1},m_{2},m_{3}).
\end{equation}
We define the covariant derivative $\mathcal{D}$ and the exterior
covariant derivative $D$ by 

\begin{equation}
\mathcal{D}=\partial_{u}+\mathrm{ad}_{E},\quad D=(\partial_{u}+\mathrm{ad}_{E})du
\end{equation}
Correspondingly, their duals are given by 

\begin{equation}
\mathcal{D}^{*}=\partial_{u}+\mathrm{ad}_{E}^{*},\quad D^{*}=(\partial_{u}+\mathrm{ad}_{E}^{*})du
\end{equation}
where $\mathrm{ad}^{*}:\mathfrak{se}(3)\times\mathfrak{se}^{*}(3)\rightarrow\mathfrak{se}^{*}(3)$
is the coadjoint action of $\mathfrak{se}(3)$ on $\mathfrak{se}^{*}(3)$,
defined via the relation

\begin{equation}
\langle\mathrm{ad}_{X}^{*}\sigma,Y\rangle=-\langle\sigma,\mathrm{ad}_{X}Y\rangle.
\end{equation}
Using this construction, the balance law takes the following equivalent
representations 
\begin{equation}
D^{*}\underline{\Sigma}+\underline{j}\,du=0,\quad\mathcal{D}^{*}\underline{\Sigma}+\underline{j}=0,\label{eq:cartan-dynamic}
\end{equation}

\emph{Constitutive laws. }Constitutive laws are given by a simple
invariant form specifying two functional relations for an elastic
material in a viscous fluid
\begin{equation}
\underline{\Sigma}=\underline{\Sigma}(\varphi,E),\quad\underline{j}=\underline{j}(\varphi,V).\label{eq:invariant-constitituve}
\end{equation}
The first (``rheological'') constitutive law relates the generalized
stress $\underline{\Sigma}$ to the configuration $\varphi$ and the
generalized deformation $\underline{E};$ the second (``dynamic'')
constitutive law relates the generalized force $\underline{j}$ to
the configuration $\varphi$ and the generalized velocity $\underline{V}$.
The boundary conditions imposed at the ends of the rod depend on the
constitutive choices, which will be discussed in detail in a later
section. Constitutive laws thus expressed automatically satisfy the
principle of material frame indifference. 

The Lie algebra-valued differential formulations -- of the kinematics
in Eq. (\ref{eq:cartan-structure}), dynamics in Eq. (\ref{eq:cartan-dynamic})
and constitutive relations in Eq. (\ref{eq:invariant-constitituve})
-- provide a concise, rigidity-invariant representation of the overdamped
motion of Cosserat rods and constitute the key results of this section.
To our knowledge, they provide the most succinct geometrized expression
of Cosserat rod motion available in literature. This parallels the
concision and geometric clarity achieved by expressing the Maxwell
equations in terms of differential forms, independently developed
by Kottler \citep{Kottler}, Cartan \citep{Cartan1986OnMW}, Van Dantzig
\citep{vanDantzig_1934} as well as the joint treatment of exterior
and interior calculus by Kähler \citep{Kahler1962}, putting mathematical
and conceptual simplification of Maxwell's work into a form that is
more accessible and influential. We summarize the four different representations
of configuration, kinematic and dynamic variables, their corresponding
equations, compatibility condition, and constitutive law using vectors,
moving frame, Lie group and differential form in Table \ref{tab:representations}.

\begin{table*}
\centering
\renewcommand{\arraystretch}{2.2}
\begin{tabular}{|p{2.8cm}|p{3.2cm}|p{3.2cm}|p{3.0cm}|p{2.8cm}|}
\hline 
\tcell{} & \tcell{\textbf{Vector}} & \tcell{\textbf{Moving Frame}} & \tcell{\textbf{Lie Group}} & \tcell{\textbf{Differential Form}} \tabularnewline
\hline 
\tcell{\textbf{Configuration}} & \multicolumn{1}{c|}{$\boldsymbol{r},\boldsymbol{e}_{i}$} & \multicolumn{1}{c|}{$\boldsymbol{r},\boldsymbol{e}_{i}$} & \multicolumn{1}{c|}{$\varphi$} & \multicolumn{1}{c|}{$\varphi$}\tabularnewline
\hline 
\tcell{\textbf{Deformation}} & \multicolumn{1}{c|}{$\boldsymbol{h},\boldsymbol{\Pi}$} & \multicolumn{1}{c|}{$\underline{h},\;\underline{\Pi}$} & \multicolumn{1}{c|}{\footnotesize$E=\begin{pmatrix}0 & 0\\
\underline{h} & \hat{\Pi}
\end{pmatrix}$ (turn)} & \multicolumn{1}{c|}{$Edu$}\tabularnewline
\hline 
\tcell{\textbf{Velocity}} & \multicolumn{1}{c|}{$\boldsymbol{v},\boldsymbol{\Omega}$} & \multicolumn{1}{c|}{$\underline{v},\;\underline{\Omega}$} & \multicolumn{1}{c|}{\footnotesize$V=\begin{pmatrix}0 & 0\\
\underline{v} & \hat{\Omega}
\end{pmatrix}$ (twist)} & \multicolumn{1}{c|}{$Vdt$}\tabularnewline
\hline 
\tcell{\textbf{Kinematics}} & \multicolumn{1}{c|}{\scriptsize$\begin{aligned} & \boldsymbol{r}^{\prime}=\boldsymbol{h},\,\boldsymbol{e}_{i}^{\prime}=\bm{\Pi}\times\boldsymbol{e}_{i}\\
 & \dot{\boldsymbol{r}}=\boldsymbol{v},\,\dot{\boldsymbol{e}}_{i}=\bm{\Omega}\times\boldsymbol{e}_{i}
\end{aligned}$} & \multicolumn{1}{c|}{\scriptsize$\begin{aligned} & \boldsymbol{r}^{\prime}=h_{i}\boldsymbol{e}_{i},\,\boldsymbol{e}_{i}^{\prime}=\epsilon_{ijk}\Pi_{j}\boldsymbol{e}_{k}\\
 & \dot{\boldsymbol{r}}=v_{i}\boldsymbol{e}_{i},\,\dot{\boldsymbol{e}}_{i}=\epsilon_{ijk}\Omega_{j}\boldsymbol{e}_{k}
\end{aligned}$} & \multicolumn{1}{c|}{\small$\begin{aligned} & \varphi^{\prime}=\varphi E\\
 & \dot{\varphi}=\varphi V
\end{aligned}$} & \multicolumn{1}{c|}{\small$\begin{aligned} & d\varphi=\varphi\xi\\
 & \xi=Edu+Vdt
\end{aligned}$}\tabularnewline
\hline 
\tcell{\textbf{Compatibility}} & \multicolumn{1}{c|}{\scriptsize$\begin{aligned}\dot{\boldsymbol{h}}-\boldsymbol{v}^{\prime}&=0\\
\dot{\bm{\Pi}}-\bm{\Omega}^{\prime}&=\bm{\Omega}\times\bm{\Pi}
\end{aligned}$} & \multicolumn{1}{c|}{\scriptsize$\begin{aligned}\dot{\underline{h}}-\underline{v}^{\prime}&=\underline{\Pi}\times\underline{v}-\underline{\Omega}\times\underline{h}\\
\dot{\underline{\Pi}}-\underline{\Omega}^{\prime}&=\underline{\Omega}\times\underline{\Pi}
\end{aligned}$} & \multicolumn{1}{c|}{\small$\dot{E}-V^{\prime}=[E,V]$} & \multicolumn{1}{c|}{\small$d\xi+\xi\wedge\xi=0$}\tabularnewline
\hline 
\tcell{\textbf{Stress-Moment}} & \multicolumn{1}{c|}{$\boldsymbol{F},\boldsymbol{M}$} & \multicolumn{1}{c|}{$\underline{F},\;\underline{M}$} & \multicolumn{1}{c|}{$\underline{\Sigma}$} & \multicolumn{1}{c|}{$\underline{\Sigma}$}\tabularnewline
\hline 
\tcell{\textbf{Force-Torque}} & \multicolumn{1}{c|}{$\boldsymbol{j},\boldsymbol{m}$} & \multicolumn{1}{c|}{$\underline{f},\;\underline{m}$} & \multicolumn{1}{c|}{$\underline{j}$ (wrench)} & \multicolumn{1}{c|}{$\underline{j}du$}\tabularnewline
\hline 
\tcell{\textbf{Balance law}} & \multicolumn{1}{c|}{\scriptsize$\begin{aligned} & \boldsymbol{F}^{\prime}+\boldsymbol{f}=0\\
 & \boldsymbol{M}^{\prime}+\boldsymbol{h}\times\boldsymbol{F}+\boldsymbol{m}=0
\end{aligned}$} & \multicolumn{1}{c|}{\tiny$\begin{aligned} & \underline{F}^{\prime}+\underline{\Pi}\times\underline{F}+\underline{f}=0\\
 & \underline{M}^{\prime}+\underline{\Pi}\times\underline{M}+\underline{h}\times\underline{F}+\underline{m}=0
\end{aligned}$} & \multicolumn{1}{c|}{\footnotesize$\mathcal{D}^{*}\underline{\Sigma}+\underline{j}=0$} & \multicolumn{1}{c|}{\footnotesize$D^{\star}\underline{\Sigma}+\underline{j}du=0$}\tabularnewline
\hline 
\tcell{\textbf{Constitutive law}} & \multicolumn{1}{c|}{\scriptsize$\begin{aligned} & \boldsymbol{F}=\boldsymbol{F}(\boldsymbol{h},\boldsymbol{\Pi})\\
 & \boldsymbol{M}=\boldsymbol{M}(\boldsymbol{h},\boldsymbol{\Pi})
\end{aligned}$} & \multicolumn{1}{c|}{\scriptsize$\begin{aligned} & \underline{F}=\underline{F}(\underline{h},\underline{\Pi})\\
 & \underline{M}=\underline{M}(\underline{h},\underline{\Pi})
\end{aligned}$} & \multicolumn{1}{c|}{\small$\underline{\Sigma}=\underline{\Sigma}(E)$} & \multicolumn{1}{c|}{\small$\underline{\Sigma}=\underline{\Sigma}(E)$}\tabularnewline
\hline
\end{tabular}

\caption{The table displays four different representations, namely, vectorial, moving frame, Lie group, and differential form formulations of the kinematics, dynamics and constitutive law of the Cosserat rod.\label{tab:representations}}
\end{table*}

\subsection{Elastohydrodynamics}

Since our focus is the overdamped motion of the rod, we assume that
the dissipative part of the generalized body force density is proportional
to the generalized velocity,
\begin{equation}
\underline{j(}\varphi,V)=-\underline{\underline{\Gamma}}\cdot\underline{V}+\underline{j}(\varphi)
\end{equation}
where $\underline{j}(\varphi)$ denotes non-dissipative components
and the rigidity-invariant Stokes friction matrix $\underline{\underline{\Gamma}}$
is symmetric and positive-definite. With these constitutive choices,
the overdamped balance law takes the new form
\begin{equation}
\underline{\underline{\Gamma}}\cdot\underline{V}=\mathcal{D}^{*}\underline{\Sigma}(E)+\underline{j}(\varphi),
\end{equation}
and the generalized velocity is \emph{instantaneously} determined
by the generalized deformation via the relation
\begin{equation}
\underline{V}(\varphi)=\underline{\underline{\Gamma}}^{-1}[\mathcal{D}^{*}\underline{\Sigma}(E)+\underline{j}(\varphi)]\label{eq:instantaneously-determined-velocity}
\end{equation}
Eliminating velocity between the kinematic and dynamic equations,
and substituting the generalized deformation using Eq. (\ref{eq:kinematics-group}),
a closed system of 12 non-linear partial differential equations for
the components of the $SE(3)$-valued matrix function $\varphi(u,t)$
\begin{equation}
\dot{\varphi}=\varphi V(\varphi),\quad\underline{V}=\underline{\underline{\Gamma}}^{-1}[\mathcal{D}^{*}\underline{\Sigma}(\varphi^{-1}\varphi^{\prime})+\underline{j}(\varphi)]\label{eq:geometrised-eqn}
\end{equation}
Due to the overdamped approximation, the spatial kinematic equation
is naturally exploited in the construction. As the generalized velocity
is determined by the balance of generalized stresses and forces, this
one equation integrates both the kinematic and dynamic content of
Cosserat rod theory.

Given a choice of local coordinates of the Lie group Eq. (\ref{eq:geometrised-eqn})
can be transformed into a set of 6 equations, corresponding to each
dimension of $SE(3)$. Such a reformulation would enjoy the benefit
of being geometrically exact and structure-preserving of the Lie group,
which is of importance in numerical simulations. Such a procedure
will be demonstrated in Sec.~\ref{sec:Coordinatised-equations}.

Lastly, we should note that in the case of purely constitutive dynamics
(that is, if there is no $\varphi$-dependence in Eq. (\ref{eq:invariant-constitituve})),
then it is possible to arrive at a simplified and equivalent form
of the equations of motion, entirely in terms of Lie algebraic quantities.
Using Eq. (\ref{eq:compatibility-condition-group}) and Eq. (\ref{eq:instantaneously-determined-velocity})
we can write

\begin{equation}
\dot{E}=\partial_{u}V+[E,V],\quad V=\underline{\underline{\Gamma}}^{-1}[\mathcal{D}^{*}\underline{\Sigma}(E)].\label{eq:geometrised-eqn-lie}
\end{equation}
These are a closed system of 6 non-linear partial differential equations,
corresponding to each degree of freedom in the deformation field in
Eq. (\ref{eq:E}). The solutions to Eq. (\ref{eq:geometrised-eqn-lie})
can be used with Eq. (\ref{eq:kinematics-group}) to reconstruct the
full temporal evolution of the Cosserat rod configuration. As these
equations are defined on the vector space $\mathfrak{se}(3)$, as
opposed to the manifold $SE(3)$, they are by construction left-invariant,
geometrically exact and structure-preserving.

\section{\label{sec:Constitutive-laws}Work, Constitutive laws and boundary
conditions}

Constitutive laws provide the essential link between kinematics and
dynamics by relating stresses to deformations. For Cosserat rods,
these relationships should account for the full range of deformations
-- stretching, shearing, twisting and bending -- while respecting
material symmetries and frame indifference. A fundamental question
is: when do these constitutive relations derive from an energy principle,
and when do they represent dissipative or active processes? Multiplying
the first equation by the linear velocity $\underline{v}$ and the
second by the angular velocity $\underline{\Omega}$, and integrating
over space, we have

\begin{widetext}
\begin{align}
\int_{0}^{L}(\underline{F}^{\prime}+\underline{\Pi}\times\underline{F}+\underline{f})\cdot\underline{v}+(M^{\prime}+\underline{\Pi}\times\underline{M}+\underline{h}\times\underline{F}+\underline{m})\cdot\underline{\Omega}\;du=0.
\end{align}

\end{widetext}

Reformulating the integrands in the Lie group shows
\begin{align}
\int_{0}^{L}\big(\underline{\Sigma}^{\prime}+\mathrm{ad}_{E}^{*}\underline{\Sigma}+\underline{j}\big)\underline{V}\;du=0.
\end{align}

Integrating by parts on $u$ and using the definition of coadjoint
action $\mathrm{ad}^{*}:\mathfrak{se}(3)\rightarrow\mathfrak{se}^{*}(3)$
and the compatibility condition $\dot{E}-V^{\prime}=\mathrm{ad}_{E}V$,
the integral becomes
\begin{align}
\int_{0}^{L}\big(-\langle\Sigma,\dot{E}\rangle+\langle j,V\rangle\big)\;du+\langle\Sigma,V\rangle\biggr|_{0}^{L}=0.\label{eq:energy_lie_int}
\end{align}
For a constitutive generalized stress $\Sigma=\Sigma\left(E\right)$
and external force $j=j\left(\varphi\right)$ to be conservative,
they each need to satisfy certain potential conditions in the work-energy
principle sense. To derive these conditions, we demand the integral
in Eq. (\ref{eq:energy_lie_int}) to be minus the time derivative
of some stored energy functional $\mathcal{E}$, which we assume to
be a functional of the deformation $E$ and the configuration $\varphi$
in the form of a sum of the elastic energy and potential energy functionals
\[
\mathcal{E}\left[E,\varphi\right]=\mathcal{E}_{\text{el}}\left[E\right]+\mathcal{E}_{\text{pot}}\left[\varphi\right]=\int\left(\mathcal{\varepsilon}_{\mathrm{el}}\left(E\right)+\mathcal{\varepsilon}_{\mathrm{pot}}\left(\varphi\right)\right)du.
\]
The rate of change of stored energy is $\dot{\mathcal{E}}=\dot{\mathcal{E}}_{\text{el}}+\dot{\mathcal{E}}_{\text{pot}}$,
where
\begin{equation}
\dot{\mathcal{E}}_{\text{el}}=\int\left\langle \frac{\partial\varepsilon_{\text{el}}}{\partial E},\dot{E}\right\rangle du,\quad\dot{\mathcal{E}}_{\text{pot}}=\int\frac{\partial\varepsilon_{\text{pot}}}{\partial\varphi_{ij}}\dot{\varphi}_{ij}du.\label{eq:en_rate_change}
\end{equation}
Comparing Eq. (\ref{eq:en_rate_change}) with (\ref{eq:energy_lie_int})
gives us immediately an expression for conservative elastic stresses
as $\underline{\Sigma}={\displaystyle \frac{\partial\mathcal{\varepsilon}_{\text{el}}}{\partial\underline{E}}}.$
A necessary condition for the conservativeness of constitutive stresses
then derives from the equality of mixed partial derivatives and reads
as
\begin{equation}
\frac{\partial\Sigma_{i}}{\partial E_{j}}=\frac{\partial\Sigma_{j}}{\partial E_{i}}.\label{eq:int_stress}
\end{equation}
The integrability conditions for constitutive forces are somewhat
more involved. From the kinematic equation (\ref{eq:kinematics-group})
and Eq. (\ref{eq:en_rate_change}), we get (repeated indices are summed
over)
\begin{equation}
\dot{\mathcal{E}}_{\text{pot}}=\int\frac{\partial\varepsilon_{\text{pot}}}{\partial\varphi_{ij}}\varphi_{ik}V_{kj}du=\int\frac{\partial\varepsilon_{\text{pot}}}{\partial\varphi_{ij}}\varphi_{ik}\left(b_{\alpha}\right)_{kj}V_{\alpha},\label{eq:epot_expanded}
\end{equation}
where in the second equality we expanded the Lie algebra element $V$
in the basis $b_{\alpha}$. Comparing Eq. (\ref{eq:energy_lie_int})
with (\ref{eq:epot_expanded}), we find the following expression for
the forces
\begin{equation}
j_{\alpha}=-\varphi_{ik}\left(b_{\alpha}\right){}_{kj}\frac{\partial\mathcal{\varepsilon}_{\text{pot}}}{\partial\varphi_{ij}}\equiv-X_{\alpha}\left[\varepsilon_{\text{pot}}\right],\label{eq:j_vec_field}
\end{equation}
where we have introduced the differential operator 
\[
X_{\alpha}=\varphi_{ik}\left(b_{\alpha}\right){}_{kj}\frac{\partial}{\partial\varphi_{ij}}
\]
on the Lie group. The $X_{\alpha}$ are left-invariant vector fields
on the Lie group associated with the Lie algebra basis $b_{\alpha}$
and satisfy the commutation relations $\left[X_{\alpha},X_{\beta}\right]=C_{\alpha\beta\gamma}X_{\gamma}$.They
generate infinitesimal translations and rotations along the moving
frame vectors. The integrability conditions on the constitutive forces
$j_{\alpha}$ follow from the antisymmetric property of the Lie bracket
of the left-invariance of vector fields in Eq. (\ref{eq:j_vec_field}):
\begin{equation}
X_{\alpha}\left[j_{\beta}\right]-X_{\beta}\left[j_{\alpha}\right]=C_{\alpha\beta\gamma}j_{\gamma}.
\end{equation}
A version of this integrability condition for the motion of a particle
on a manifold has been obtained previously \citep{PhysRevE.111.045418}
. Choosing the standard basis of $\mathfrak{se}\left(3\right)$ for
$b_{\alpha}$, we find that the translational left-invariant vector
fields are given by
\[
X_{1}=\boldsymbol{e}_{1}\cdot\frac{\partial}{\partial\boldsymbol{r}},\quad X_{2}=\boldsymbol{e}_{2}\cdot\frac{\partial}{\partial\boldsymbol{r}},\quad X_{3}=\boldsymbol{e}_{3}\cdot\frac{\partial}{\partial\boldsymbol{r}},
\]
while the rotational ones are
\begin{align*}
X_{4} & =\boldsymbol{e}_{2}\cdot\frac{\partial}{\partial\boldsymbol{e}_{3}}-\boldsymbol{e}_{3}\cdot\frac{\partial}{\partial\boldsymbol{e}_{2}},\\
X_{5} & =\boldsymbol{e}_{3}\cdot\frac{\partial}{\partial\boldsymbol{e}_{1}}-\boldsymbol{e}_{1}\cdot\frac{\partial}{\partial\boldsymbol{e}_{3}},\\
X_{6} & =\boldsymbol{e}_{1}\cdot\frac{\partial}{\partial\boldsymbol{e}_{2}}-\boldsymbol{e}_{2}\cdot\frac{\partial}{\partial\boldsymbol{e}_{1}}.
\end{align*}

For simplicity, we assume the following linear constitutive law, stating
that the generalized stress is proportional to the generalized strains,
\begin{align}
\underline{\Sigma}(\underline{E})=\underline{\underline{K}}(\underline{E}-\underline{\bar{E}}).\label{eq:general-constitutive-law}
\end{align}
Here, the rigidity-invariant Hookean stiffness matrix $\underline{\underline{K}}$
is positive definite\textcolor{black}{, $\underline{\bar{E}}$ is
the deformation in the reference state and $\underline{E}-\underline{\bar{E}}$
is the strain in the deformed state. }We note that this only requires
the material\emph{ }response to be \emph{mechanically} linear but
places no restriction on the configurational motion, which admits
large configurational changes and remains fully \emph{geometrically}
non-linear. This ensures that the motions are not restricted in configuration
space. Going forward, for simplicity, we will drop the underline notations
denoting vector-valued quantities, but it shouldn't be confused with
matrix-valued Lie algebra elements. 

Assuming a diagonal stiffness matrix $\underline{\underline{K}}=\text{\text{diag}}(k_{1},k_{2},k_{3})$,
where \textcolor{black}{the material parameters $k_{1},k_{2},k_{3}$
represent the stretching, shearing and rotational elastic coefficients},
the components of the constitutive relations (\ref{eq:general-constitutive-law})
read
\begin{equation}
F_{1}=k_{1}(h_{1}-1),\quad F_{2}=k_{2}h_{2},\quad M=k_{3}\Pi.\label{eq:hyperelastic-constitutive-laws}
\end{equation}
This constitutive law is hyperelastic and is suited for Cosserat elasticity.
The energy functional in this case is
\begin{equation}
\mathcal{\mathcal{U}}=\dfrac{1}{2}\int k_{1}(h_{1}-1)^{2}+k_{2}h_{2}^{2}+k_{3}\Pi^{2}\;du,\label{eq:energy-functional}
\end{equation}
encoding the energetic penalty for a deformation from the reference
state $\bar{E}_{}=(1,0,0)$ corresponding to a planar rod with no
stretching $(h_{1}=1)$, no shearing $(h_{2}=0)$ and no bending $(\Pi=0)$.

The corresponding boundary conditions for our system of interest are
\textit{clamped}\textbf{ }and \textit{free} boundary conditions at
two ends of the rod, respectively.  In geometric form, the clamped
boundary condition is
\begin{equation}
\varphi=\varphi_{0}=\mathrm{const}.,
\end{equation}
while the free boundary condition is
\begin{equation}
\Sigma=0
\end{equation}

\section{\label{sec:Linearisation}Linearization}

The geometric field theory formulation reveals the rich structure
in Cosserat rod dynamics, but extracting physical insights from the
nonlinear $SE(3)$-valued PDE system remains challenging. Linearization
about an equilibrium configuration provides a systematic approach
to analyze stability, identify dominant physical mechanisms, and validate
numerical schemes against analytical benchmarks. Consider a static
equilibrium solution $\varphi_{0}(u)$ satisfying the overdamped equations
of motion with vanishing generalized velocity,$V_{0}=0$. Let $\epsilon\ll1$
be a small perturbation parameter. We linearize about the base state
in the Lie group by writing
\begin{equation}
\begin{aligned}\varphi & =\varphi_{0}\exp(\epsilon\eta)\\
 & =\varphi_{0}(I+\epsilon\eta)+\mathcal{O}(\epsilon^{2})
\end{aligned}
\end{equation}
where $\eta\in\mathfrak{se}(3)$ is a Lie algebra element we refer
to as the displacement field. Correspondingly, expanding the relations
$E=\varphi^{-1}\varphi^{\prime}$, $V=\varphi^{-1}\dot{\varphi}$
gives
\begin{equation}
\begin{aligned}V & =\epsilon\delta V+\mathcal{O}(\epsilon^{2})\\
E & =E_{0}+\epsilon\delta E+\mathcal{O}(\epsilon^{2})
\end{aligned}
\end{equation}
where
\begin{align}
\delta V & =\dot{\eta}\\
\delta E & =\eta^{\prime}+[E_{0},\eta]\label{eq:linearised-strain}
\end{align}

Then, the stress $\Sigma=\Sigma(E)$ and external force $j=j(\varphi)$
are linearized as
\begin{align}
\delta\Sigma=T\delta E,\quad\delta j=S\eta,
\end{align}

where, by definitions,
\begin{align}
T:=\frac{\partial\Sigma}{\partial E}\biggr|_{E=E_{0}}\quad\text{and}\quad S:=\frac{\partial j}{\partial\varphi}\biggr|_{\varphi=\varphi_{0}}\varphi_{0}
\end{align}

are the Jacobian matrices. 

Finally, we linearize the closed geometrized equation $(\ref{eq:geometrised-eqn})$,
in which the linearization of $\dot{\varphi}=\varphi V$ yields $\dot{\eta}=\delta V$
while linearization of the velocity expression $\underline{V}=\underline{\underline{\Gamma^{-1}}}[\mathcal{D}^{*}\underline{\Sigma}(\varphi^{-1}\varphi^{\prime})+\underline{j}(\varphi)]$
is calculated by
\begin{align}
\begin{split}V=\delta V & =\Gamma^{-1}\big[\mathcal{D}^{*}(T\delta E)+S\eta\big]\\
 & =\Gamma^{-1}\big[\mathcal{D}^{*}\big(T(\eta^{\prime}+\mathrm{ad}_{E_{0}}\eta)\big)+S\eta\big],
\end{split}
\end{align}

using the linearization of strain $(\ref{eq:linearised-strain})$
in the last line.

Putting these all together, we arrive at the linearized geometrized
equation of motion
\begin{align}
\dot{\eta}=\Gamma^{-1}\mathcal{D}^{*}T\eta{}^{\prime}+\Gamma^{-1}\big[\mathcal{D}^{*}T(\mathrm{ad}_{E_{0}})+S\big]\eta.
\end{align}

This can be further cast into a diffusion-advection-reaction form
$\dot{\eta}=D\eta^{\prime\prime}+A\eta^{\prime}+R\eta$, by collecting
terms with $\eta^{\prime\prime},\eta^{\prime},\eta$, as

\begin{widetext}
\begin{align}
\dot{\eta}=(\Gamma^{-1}T)\eta^{\prime\prime}+\Gamma^{-1}\big(T^{\prime}-\mathrm{ad}_{E}T+T\mathrm{ad}_{E_{0}}\big)\eta^{\prime}+\Gamma^{-1}\big(T\mathrm{ad}_{E_{0}^{\prime}}+T^{\prime}\mathrm{ad}_{E_{0}}-(\mathrm{ad}_{E}T\mathrm{ad}_{E_{0}})+S\big)\eta.
\end{align}
\end{widetext}

\section{\label{sec:Coordinatised-equations}Coordinatized equations }

The geometrized equations of motion can be cast entirely in terms
of the configuration as a set of partial differential equations on
the Lie group, where the generalized deformation and its derivative
have been eliminated in favor of derivatives of the configuration,
as shown in Eq. (\ref{eq:geometrised-eqn}). This invariant equation
admits expressions in local coordinates on $SE(3)$, and any valid
local coordinate system may be used. Since there are least seven coordinate
systems for rotations in common use (Euler angles, orthogonal matrices,
quaternions, direction cosines, Cayley-Klein parameters, Gibbs vectors,
and bivectors) our geometrized equations are valuable for expressing
their coordinate-invariant content. We illustrate how coordinate-specified
equations of motion can be obtained once a choice of local coordinates
has been made. For simplicity, we demonstrate this for planar Cosserat
rod motion, where the configuration is the matrix Lie group $SE(2)$,
the special Euclidean group of proper 2D rigid motions in Euclidean
space $\mathbb{E}^{2}$, and the coordinatization comprises of 2D
translation $x,y$ and rotation specified by the angle $\theta$ between
the moving frame and the $x$-axis. 

Confining motion to the $x\text{{-}}y$ plane, the centerline is coordinatized
as $\boldsymbol{r}=(x,y,0)$ and the three frame vectors as $\boldsymbol{e}_{1}=(\cos\theta,\sin\theta,0)$,
$\boldsymbol{e}_{2}=(-\sin\theta,\cos\theta,0)$, $\boldsymbol{e}_{3}=(0,0,1$).
The configuration is then specified by a $3\times3$ matrix in $SE(2)$,
\begin{equation}
\varphi=\left[\begin{array}{ccc}
1 & 0 & 0\\
x & \cos\theta & -\sin\theta\\
y & \sin\theta & \cos\theta
\end{array}\right].
\end{equation}
The generalized deformation and generalized velocity are given by
$3\times3$ matrices in the Lie algebra $\mathfrak{se}(2),$
\begin{equation}
E=\left[\begin{array}{ccc}
0 & 0 & 0\\
h_{1} & 0 & -\Pi\\
h_{2} & \Pi & 0
\end{array}\right],\quad V=\left[\begin{array}{ccc}
0 & 0 & 0\\
v_{1} & 0 & -\Omega\\
v_{2} & \Omega & 0
\end{array}\right],
\end{equation}
where the Lie-algebra valued quantities are made up of $2\times1$
vectors $\underline{h}=(h_{1},h_{2})^{T}$, $\underline{v}=(v_{1},v_{2})^{T}$
and $2\times2$ antisymmetric matrices $\hat{\Pi},\hat{\Omega}$.
The corresponding vectors are $\underline{E}=(h_{1},h_{2},\Pi)$ and
$\underline{V}=(v_{1},v_{2},\Omega).$ In coordinates, the spatial
kinematic equation $\varphi^{\prime}=\varphi E$ for the generalized
deformation can be decomposed into three equations
\begin{align}
\begin{split}x^{\prime} & =h_{1}\cos\theta-h_{2}\sin\theta,\\
y^{\prime} & =h_{1}\sin\theta+h_{2}\cos\theta,\\
\theta^{\prime} & =\Pi,
\end{split}
\label{eq:coordinate-strain}
\end{align}
Correspondingly, the temporal kinematic equation $\dot{\varphi}=\varphi V$
for the generalized velocity has the three components as follows
\begin{align}
\begin{split}\dot{x} & =v_{1}\cos\theta-v_{2}\sin\theta,\\
\dot{y} & =v_{1}\sin\theta+v_{2}\cos\theta,\\
\dot{\theta} & =\Omega,
\end{split}
\label{eq:coordinate-velocity}
\end{align}
The compatibility condition $\dot{E}=V^{\prime}+[E,V]$ in components
reads
\begin{align}
\begin{split}\dot{h}_{1} & -v_{1}^{\prime}=\Omega h_{2}-\Pi v_{2},\\
\dot{h}_{2} & -v_{2}^{\prime}=\Omega h_{1}-\Pi v_{1},\\
\dot{\Pi} & =\Omega^{\prime},
\end{split}
\label{eq:compatibility-conditions-1}
\end{align}
which gives the compatible time evolution of the generalized deformation
in terms of the generalized velocity and its spatial derivatives.
The generalized stress has components $\underline{\Sigma}=(F_{1,},F_{2},M)$.
Assuming a diagonal friction matrix $\Gamma=\text{diag}(\gamma_{1},\gamma_{2},\gamma_{3})$
, the components of the balance law $\Gamma\underline{V}=\mathcal{D}^{*}\underline{\Sigma}+\underline{j}$
are given by
\begin{align}
\begin{split}\gamma_{1}v_{1} & =F_{1}^{\prime}-\Pi F_{2}+f_{1},\\
\gamma_{2}v_{2} & =F_{2}^{\prime}+\Pi F_{1}+f_{2},\\
\gamma_{3}\Omega & =M^{\prime}+h_{1}F_{2}-h_{2}F_{1}+m.
\end{split}
\label{eq:balance-laws}
\end{align}

The equations are closed when constitutive choices are made, relating
stress and deformation. At the same time, the velocity is eliminated
between the above balance laws and the temporal kinematic equation,
providing the coordinatized equations of motion,

\begin{widetext}
\begin{align}
\begin{split}\dot{x} & =\frac{1}{\gamma_{1}}(F_{1}'-\theta'F_{2})\cos\theta-\frac{1}{\gamma_{2}}(F_{2}'+\theta'F_{1})\sin\theta\\
\dot{y} & =\frac{1}{\gamma_{1}}(F_{1}'-\theta'F_{2})\sin\theta+\frac{1}{\gamma_{2}}(F_{2}'+\theta'F_{1})\cos\theta\\
\dot{\theta} & =\frac{1}{\gamma_{3}}\big[M'+\big(x'\sin\theta-y'\cos\theta\big)F_{1}+\big(x'\cos\theta+y'\sin\theta\big)F_{2}\big]
\end{split}
\label{eq:coordinatised-eqns-1}
\end{align}

\end{widetext}

\textcolor{black}{}

where $\gamma_{1},\gamma_{2},\gamma_{3}$ represent the stretching,
shearing and rotational friction coefficients. The clamped and free
boundary conditions in their general forms correspond to
\begin{align}
\varphi=\mathrm{const}
\end{align}

and
\begin{align}
\Sigma=0\quad\iff\quad F_{1}=0,\;F_{2}=0,\;M=0,
\end{align}

at the endpoints, respectively.

Now, under the hyperelastic constitutive laws $F_{1}=k_{1}(h_{1}-1),F_{2}=k_{2}h_{2},M=k_{3}\Pi$
with $\gamma_{1}=\gamma_{2}=\gamma$ and $k_{1}=k_{2}=k$, the above
nonlinear coordinatized equations of motion simplify to 
\begin{align}
\gamma\dot{x} & =k\left(x''+\,\theta'\sin\theta\right),\nonumber \\
\dot{\gamma y} & =k\left(y''-\,\theta'\cos\theta\right),\nonumber \\
\gamma_{3}\dot{\theta} & =k_{3}\theta''+k\left(y'\cos\theta\,-\,x'\sin\theta\right)
\end{align}

with corresponding clamped-free boundary conditions in components
\begin{align}
x=0,\quad y=0,\quad\theta=0
\end{align}

and
\begin{equation}
\begin{split}x^{\prime}\cos\theta+y'\sin\theta & =\bar{h}_{1}\\
-x'\sin\theta+y'\cos\theta & =\bar{h}_{2}\\
\theta' & =\bar{\Pi}
\end{split}
\end{equation}

These are a system of three coupled parabolic non-linear partial differential
equations for the coordinates $x(u,t)$, $y(u,t)$ and $\theta(u,t)$
that determine the overdamped planar motion of a materially linear
rod that can stretch, shear and bend but not twist. They are valid
for arbitrarily large deformations and are geometrically exact.

\section{\label{sec:Beam-limits}Beam limits}

So far, we have demonstrated that the geometric formulation encompasses
both nonlinear and linear regimes within a unified framework. The
latter contains a greater amount of information tied closely to some
well-known beam theories in classical elasticity. Under certain constraints,
our linearized system naturally reduces to several familiar beam equations,
showing that classical theory emerges as a special case rather than
a separate formulation. We will show that in the no stretching and
no shearing limits our Cosserat rod model reduces to the Euler-Bernoulli
beam theory \citep{Timoshenko1983,BauchauCraig,Hodges2006}, which
considers an elastic rod that can only bend in one plane. 

The vanishing of stretching and shearing is equivalent to freezing
the kinematic variables, that is, $h_{1}=1,h_{2}=0.$ Using these
in the balance laws $(\ref{eq:balance-laws})$ and the compatibility
conditions $(\ref{eq:compatibility-conditions-1})$ with the assumption
that the angle relaxes fast to its equilibrium, we obtain
\begin{align}
M^{\prime}+F_{2}=0,
\end{align}
and
\begin{align}
0 & =v_{1}{}^{\prime}-\Pi v_{2}\\
0 & =v_{2}{}^{\prime}+\Pi v_{1}-\Omega.
\end{align}

Then, substituting the compatibility conditions into the first two
balance law equations
\begin{align}
\begin{split}\gamma_{1}v_{1} & =F_{1}{}^{\prime}+\Pi M^{\prime}\\
\gamma_{2}v_{2} & =-M^{\prime\prime}+\Pi F_{1}
\end{split}
\end{align}

yields
\begin{align}
0 & =\frac{1}{\gamma_{1}}\big(F_{1}{}^{\prime\prime}+(\Pi M^{\prime})^{\prime}\big)-\frac{1}{\gamma_{2}}\Pi(-M^{\prime\prime}+\Pi F_{1})\\
\Omega & =\frac{1}{\gamma_{2}}\big(-M^{\prime\prime}+\Pi F_{1}\big)^{\prime}+\frac{1}{\gamma_{1}}\Pi\big(F_{1}{}^{\prime}+\Pi M^{\prime}\big)
\end{align}

Since $\Omega=\dot{\theta},\Pi=\theta^{\prime}$, these two equations
become equations of $\theta$ and $F_{1}$, as
\begin{align}
\begin{split}0 & =\frac{1}{\gamma_{1}}\big[F_{1}{}^{\prime\prime}+(\theta^{\prime}M^{\prime})^{\prime}\big]-\frac{1}{\gamma_{2}}\big(-\theta^{\prime}M^{\prime\prime}+(\theta^{\prime})^{2}F_{1}\big)\\
\dot{\theta} & =\frac{1}{\gamma_{2}}\big(-M^{\prime\prime}+\theta^{\prime}F_{1}\big)^{\prime}+\frac{1}{\gamma_{1}}\big(\theta^{\prime}F_{1}{}^{\prime}+(\theta^{\prime})^{2}M^{\prime}\big).
\end{split}
\end{align}
Assuming the constitutive law $M=k_{3}\Pi=k_{3}\theta^{\prime}$,
the above equations further reduce to nonlinear Euler-Bernoulli beam
equations

\begin{widetext}
\begin{align}
\begin{split}0 & =\frac{1}{\gamma_{1}}\big(F_{1}{}^{\prime\prime}+k_{3}(\theta^{\prime\prime})^{2}+k_{3}\theta^{\prime}\,\theta^{\prime\prime\prime}\big)-\frac{1}{\gamma_{2}}\big(-k_{3}\theta^{\prime}\,\theta^{\prime\prime\prime}+(\theta^{\prime})^{2}F_{1}\big)\\
\dot{\theta} & =\frac{1}{\gamma_{2}}\big(-k_{3}\theta^{\prime\prime\prime\prime}+\theta^{\prime\prime}F_{1}+\theta^{\prime}F_{1}{}^{\prime}\big)+\frac{1}{\gamma_{1}}\big(\theta^{\prime}F_{1}{}^{\prime}+k_{3}(\theta^{\prime})^{2}\theta^{\prime\prime}\big).
\end{split}
\end{align}
\end{widetext}

where $F_{1}$ plays the role of a Lagrange multiplier enforcing inextensibility
and $\theta$ is the only degree of freedom. The linearized equations
capture the key structure of Euler-Bernoulli equations, as
\begin{align}
0 & =\frac{1}{\gamma_{1}}F_{1}{}^{\prime\prime}+\frac{k_{3}}{\gamma_{1}}(\theta^{\prime\prime})^{2}\\
\dot{\theta} & =-\frac{k_{3}}{\gamma_{2}}\theta^{\prime\prime\prime\prime}.
\end{align}

\section{\label{sec:Discussion-and-Conclusion}Discussion and Conclusion}

In this work, we have developed a systematic geometrization of Cosserat
rod theory in viscous environments, advancing from vectorial formulations
through moving frames to a coordinate-free field theory on $SE(3)$.
This geometric framework offers distinct advantages over classical
formulations (e.g. \citep{Antman,Love1927,Kirchhoff1859,Clebsch1883})
and modern vectorial approaches (e.g. \citep{Reissner1972,Simo1985}).

Our formulation yields coordinate-independent equations invariant
under rigid motions, naturally incorporating material frame indifference
and enabling constitutive modeling through group symmetries. The structure
unifies kinematics, dynamics, and constitutive laws into compact differential-form
expressions, $d\varphi=\varphi\xi$ and $\mathcal{D}^{*}\Sigma+jdu=0$,
which preserve the underlying $SE(3)$ geometry. In viscous regimes,
inertial terms vanish, producing a field theory distinct from prior
geometric mechanics focused on underdamped Hamiltonian systems. This
perspective also facilitates structure-preserving numerical schemes
that maintain invariants and exhibit high stability, which will be
addressed in future work.

The coordinate-free setting avoids singularities inherent in Euler-angle-based
methods and provides a systematic pathway for constitutive modeling.
Classical beam theory emerges as the small-deformation limit, validating
our approach while clarifying the origin of nonlinear effects from
the full $SE(3)$ structure. While we illustrated the formulation
in $SE(2)$, it generalizes naturally to $SE(3)$ through appropriate
retraction maps. Beyond clamped and free boundaries, pressure-driven
conditions, such as follower forces, can be incorporated. Similarly,
elastic responses that go beyond hyperelasticity, to odd and non-reciprocal
elasticity, can be incorporated. These points are reserved for future
work.

This framework establishes the foundation for geometric discretization
schemes in computational rod mechanics, including 1D discrete rods
and 2D discrete shells , in line with developments in discrete differential
geometry and structure-preserving algorithms. More broadly, it enables
systematic treatment of active materials and non-conservative forces,
with potential applications in soft robotic actuation. Thus, Cosserat
rod theory emerges as a natural geometric field theory, revealing
deep links between continuum mechanics and differential geometry while
offering practical computational advantages.
\begin{acknowledgments}
We thank Professors M. E. Cates, D. D. Holm and R. E. Goldstein for
valuable discussions. MY acknowledges the Friday afternoon Geometric
Mechanics Research Group at Imperial College London and Dr R. Hu for
several helpful remarks. RA thanks Professors A. McRobie and W. F.
Baker for insightful discussion on the engineering applications of
Cosserat theory. MY is supported by a Cambridge Commonwealth, European
and International Trust and China Scholarship Council (CSC Cambridge)
joint scholarship - CSC No. 202408060223. MW acknowledges a PhD studentship
from the EPSRC/UKRI. BN is supported by an Engineering and Physical
Sciences Research Council Grant No. EP/W524141/1 (B.N.).
\end{acknowledgments}

\bibliographystyle{apsrev}
\bibliography{cosserat-geometrisation}

\end{document}